# The Symmetries of the Order Parameters for Pairing and Phase Coherence in Hole-Doped Cuprates

A. Mourachkine

*Université Libre de Bruxelles, Service de Physique des Solides, CP233, Boulevard du Triomphe, B-1050 Brussels, Belgium*

___

**Abstract**

The symmetries of the order parameter (OP) responsible for pairing and the OP responsible for phase coherence in hole-doped cuprates are discussed. We analyze angle-resolved tunneling, ARPES, Raman scattering and torque measurements performed on hole-doped cuprates. We conclude that, most likely, the OP responsible for phase coherence has the $d_{x^2-y^2}$ symmetry whereas the OP responsible for pairing has an anisotropic s-wave symmetry.

___

## 1. Introduction

In conventional superconductors, the mechanism responsible for formation of the Cooper pairs and the mechanism responsible for establishment of the phase coherence are identical: both the phenomena occur due to phonons almost simultaneously at $T_c$. There is a consensus that, in superconducting (SC) copper-oxides, the Cooper pairs are formed in the underdoped regime well above $T_c$ [1-4]. Moreover, there are strong indications that the origins of the two mechanisms are different [3,1,5]. First of all, the magnitudes of the order parameters (OPs) responsible for pairing and phase coherence in hole-doped cuprates have different dependencies on hole concentration, $p$ in CuO$_2$ planes [1,5]. The magnitude of the OP responsible for phase coherence, $\Delta_c$, which is proportional to $T_c$, has the parabolic dependence on $p$ [1,5]. While the magnitude of the OP responsible for pairing, $\Delta_p$ increases linearly with the decrease of hole concentration [1,5,6]. Figure 1 shows the phase diagram of the two energy gaps [7] in hole-doped cuprates [1]. In Fig. 1, the $\Delta_c$ scales with $T_c$ as $2\Delta_c/k_B T_c = 5.45$ [1]. It is worth noting that both the two gaps are SC-like. Secondly, an applied magnetic field affects the two OPs in cuprates differently: the phase coherence disappears first while, in order to break the Cooper pairs, the magnitude of magnetic field must be much higher [8,3]. In general, the origin of OP defines it's symmetry. Thus, the knowledge of the symmetry of OP can help to identify it's origin. There is a general consensus that the predominant OP in hole-doped cuprates has the $d_{x^2-y^2}$ (hereafter, d-wave) symmetry [9]. This implies that one OP out of the two OPs has the d-wave symmetry. What is the symmetry of the second OP in hole-doped cuprates? For instance, in-plane torque anisotropy measurements on $Tl_2Ba_2CuO_{6+x}$ (Tl2201) show that besides the four-fold symmetry of total OP there is a two-fold component which was interpreted by the presence of s-wave OP [10]. There are strong indications that the coherent OP has the magnetic origin [11], thus, most likely, it has the d-wave symmetry [9].

The magnitude of the total OP in $Bi_2Sr_2CaCu_2O_{8+x}$ (Bi2212), which is equal, in general, to $\Delta = (\Delta_c^2 + \Delta_p^2)^{1/2}$ [12], has been measured by angle-resolved tunneling measurements [13]. The total OP observed in the experiment has an anisotropic character, however, without nodes. The symmetry of the total OP (presumably having the four-fold symmetry) resembles the $d_{xy}$ state. In tunneling measurements on $YBa_2Cu_3O_{6+x}$ (YBCO) and $La_{2-x}Sr_xCuO_4$ (LSCO), Tanaka with co-workers [14-16] obtained *similar* results. Moreover, in LSCO, zero-bias conductance peaks (ZBCP) have been observed in the direction of the minimum-gap magnitude, *i.e.* along Cu-O-Cu bonds. This fact suggests that the OP (or, at least, one OP) has the $d_{xy}$ symmetry. However, the authors [14] presented an explanation how the d-wave state is transformed on the surface into the $d_{xy}$ state. Angle-resolved photoemission (ARPES) measurements on Bi2212 show below and above $T_c$ the presence of OP having the d-wave symmetry [17]. It is interesting to note that the maximum magnitude of gap in the tunneling measurements [13] and associated in-plane anisotropy are consistent with the photoemission results [17], but the gap anisotropy patterns disagree by 45°. Electronic Raman scattering (ERS) measurements on Bi2212 show the presence of two OPs [18] with the magnitudes depending on hole concentration similar to the case of $\Delta_c$ and $\Delta_p$ (see Fig. 1). From ERS measurements on YBCO [19], Bi2212 [18,19,20] and Tl2201 [20], the OP responsible for pairing has the d-wave symmetry ($B_{1g}$ polarization) [18,19,20], and the OP responsible for phase coherence has a s-wave symmetry ($A_{1g}$ polarization) [18,20]. In ERS measurements on underdoped YBCO and Y-doped Bi2212 [21], the $B_{2g}$ polarization scales with $T_c$ in the underdoped regime. This fact implies that the OP responsible for phase coherence has the $d_{xy}$ symmetry. In very recent ERS measurements on Bi2212 [22], the OP responsible for pairing has been observed by the $A_{1g}$, $B_{1g}$ and $B_{2g}$ polarizations suggesting that the pairing OP can have a s-wave, the d-wave or $d_{xy}$ symmetry.

Thus, by summarizing the previous paragraph it is clear that there is a complete confusion in the question of symmetries of the two OPs. The purpose of this paper is to classify tunneling, ARPES and ERS data in order to find out the symmetries of the OPs responsible for pairing and phase coherence in hole-doped cuprates. In the paper, first of all, we consider all possible combinations of the symmetries of the two OPs, and, then we analyze and classify the data. In fact, most data which we will discuss here are obtained on Bi2212 cuprate since it is the most suitable cuprate for performing ARPES and tunneling measurements.

## 2. ARPES measurements on Bi2212

We start with ARPES measurements on Bi2212 since they are very consistent [17,23,24]. Recent ARPES studies show the presence only one OP in Bi2212, which has the d-wave symmetry [17,23]. It is seems that this OP is responsible for pairing since the OP has been detected below and above $T_c$ and the dependence of magnitude of this OP on hole concentration almost repeats the $\Delta_p(p)$ dependence shown in Fig. 1. At first sight, such situation may look like ARPES studies detect only the OP responsible for pairing, and, for some reasons, they are not able to observe the OP responsible for phase coherence. However, we will show in the next Section that it *may* be not the case.

We present now a short description of ARPES technique, which we will need further. Photoemission is a signal averaging experiment and provides more direct information for the valence-charge distribution [24]. ARPES data have a very good angle resolution, of ± 1°, but their energy resolution is rather moderate, of 15 meV [23]. ARPES measurements show the presence of nodes in Cu-Cu direction (*i.e.* at 45°), however, having the energy resolution of 15 meV they can only eliminate the existence of a node, but strictly speaking can not prove it. The photo-electron escape depth is very small, only of 3 Å [25]. At present there is no generally accepted model for the ARPES spectral function $A(\mathbf{k}, E)$ in copper-oxides, and it leads to uncertainty in the magnitude of energy gaps. Since the photoelectron spectral peak is quite broad, particularly in underdoped Bi2212 samples, it become customary to use the leading edge gap which is the difference between the midpoint of the leading edge and the Fermi energy. The true gap (or what is assumed to be the true gap) is larger than the leading edge gap since the linewidth and the energy resolution have to be added. In near optimally doped Bi2212, leading edge gap values are typically in the 20 - 25 meV range whereas true gaps determined by fits, are 10 meV larger [23]. In spite of all advantages of

ARPES technique, very recent results [26] suggest that ARPES data have to be considered with great care since it is possible that they reflect not only intrinsic properties of the density of states (DOS) of the quasiparticle excitations but some extrinsic effects as well [26]. These results [26] imply that (i) the gap magnitude measured by ARPES technique can be larger that the true one, and (ii) the difference between the magnitudes of the true and observed gaps is larger for poorly conducting solids than for relatively good conducting systems [26]. We will discuss these important results at the end of the next Section.

### 3. Angle-resolved tunneling measurements on Bi2212

Tunneling spectroscopy played a crucial role in the verification of the BCS theory since it probes *directly* the DOS of the quasiparticle excitations [27]. It is one of the best high energy-resolution technique ($\sim k_B T$) which is capable of detecting any gap in the quasiparticle excitation spectrum at the Fermi level $E_F$. The tunneling current probes a region of the order of the coherent length ξ [27].

Figure 2 shows angle-resolved tunneling data [13] obtained on overdoped Bi2212 single crystals with $T_c$ = 85 K ($p/p_m$ = 1.2). In Fig. 2, the maximum magnitude of tunneling gap (36.5 meV) is located in Cu-Cu direction, and the minimum magnitude (21 meV) is located in Cu-O-Cu bond directions. In tunneling measurements on YBCO and LSCO, the minimum magnitude of gap has been found also along the Cu-O-Cu bond direction [14-16]. Thus, tunneling measurements performed on the three different cuprates are consistent. In tunneling measurements on Pb-doped Bi2212 cuprate [28] in which $CuO_2$ planes are, in the first approximation, unaffected by Pb-doping, the maximum of gap magnitude located in Cu-Cu direction (see Fig. 2) remains unchanged in comparison with pure Bi2212 case while the minimum of the gap magnitude observed along Cu-O-Cu bond direction decreases proportionally to the change in $T_c$. Thus, the Pb-doping in Bi2212 affects only the OP responsible for phase coherence keeping the pairing OP unchanged. This implies that the pairing OP is presented mainly in Cu-Cu direction and the coherent OP is dominant in Cu-O-Cu bond direction.

In this Section, we will relay on the data shown in Fig. 2 in order to find out the symmetries of the $\Delta_c$ and $\Delta_p$. Our analysis will be based on the following facts: (i) the total OP consists of the two OPs, $\Delta = (\Delta_c^2 + \Delta_p^2)^{1/2}$ [12,29]; (ii) the maximum magnitudes of the two OPs at different hole concentrations are known (see Fig. 1) [1,5], and (iii) one OP has the d-wave symmetry [9]. In addition to this, we will use the principle of adjustment. On the basis of experimental results, one OP, either the $\Delta_c$, or $\Delta_p$, has the d-wave symmetry. By knowing the magnitudes of the $\Delta_c$ and $\Delta_p$ at $p/p_m = 1.2$ from Fig. 1, we find that only the coherent gap $\Delta_c(1.2) = 20$ meV may fit the data shown in Fig 2 *as the d-wave gap*. The $\Delta_p$ is too large for it, $\Delta_p(1.2) = 30 - 30.5$ meV. In Fig. 2, we present schematically the d-wave gap with the maximum magnitude of 20 meV. One should note that, in general, in two-gap scenario, the predominant character of one gap and it's magnitude do not relate to each other. Let us estimate the magnitudes of the pairing OP, $\Delta_p$, in different directions. It is obvious that the $\Delta_p$ has the maximum in Cu-Cu direction since the $\Delta_c$ has a node in this direction. Thus, $\Delta_{p,\,max} = 36.5$ meV. The minimum of the $\Delta_p$ is located in Cu-O-Cu bond directions. From $\Delta = (\Delta_c^2 + \Delta_p^2)^{1/2}$, we find that $\Delta_{p,\,min} = \pm (21^2 - 20^2)^{1/2} = \pm 6.5$ meV. So, we have two solutions for $\Delta_{p,\,min}$. Figure 3(a) shows *schematically* shapes of the two gaps with the two solutions for coherent gap: $\Delta_p$ (+) and $\Delta_p$ (-).

As we underlined above, at this moment, we consider all possible combinations of the two OPs, which we will analyze further. The $\Delta_p$ (-) solution shown in Fig. 3(a) has an extended s-wave symmetry [9] (turned by 45°). The $\Delta_p$ (+) solution has either an anisotropic s-wave or mixed (s+$d_{xy}$) symmetry. In the cases of the extended and anisotropic s-wave symmetries of the OP responsible for pairing, there is a small discrepancy between the maximum magnitudes of the $\Delta_p$ gaps shown in Fig. 3(a) (36.5 meV) and in Fig. 1 (30-30.5 meV). Figure 3(b) shows the mixed case for the $\Delta_p$ (+) solution. In this case, there is a very good agreement of the magnitudes of the two gaps in Figs. 1 and 3(b). From the point of view of tunneling measurements, the OP responsible for phase coherence has always the d-wave symmetry, but there are three possible cases for the symmetry of the OP responsible for pairing: an anisotropic and extended s-waves shown in Fig. 3(a), and the combination of the isotropic s-wave and $d_{xy}$ components shown in Fig. 3(b). It is important note that, in all three cases, the pairing OP has entirely or partially a s-wave symmetry. Thus, ARPES data and the analysis based on tunneling measurements concerning the symmetries of the two OPs contradict each other.

In fact, it is possible to explain this contradiction between ARPES and tunneling measurements. Let's assume that the OP responsible for phase coherence has the d-wave symmetry. In optimally doped and overdoped Bi2212, one should note that the magnitude of the leading edge gap at different hole concentrations coincides with the magnitude of the $\Delta_c$ shown in Fig. 1. So, in overdoped regime, probably, it is not necessary to add 10 meV to the magnitude of measured gap [23] since the same amount of photo-electron energy may originate from some extrinsic effects [26]. In the underdoped regime, according to the Joynt's results [26] the amount of photo-electron energy which originates from extrinsic effects should increase. So, the magnitude of measured gap increases with the decrease of hole concentration instead of following the parabolic dependence. In this case, it is necessary to explain how the $\Delta_c$ gap evolves into the $\Delta_p$ at $T_c$. The next question is why ARPES do not observe the second OP. Probably, it is due to a very small escape depth of photo-electrons (3 Å) [25]. In Section 5, in order to explain the discrepancy between ARPES and tunneling data we will consider the contrary assumption, namely, that the angle-resolved tunneling data are disagree with the real case by 45°.

In addition to the combinations of the two OPs presented in Fig. 3, it is possible to form, at least, one combination more. Theoretically, the d- and g-wave magnetically mediated SCs may co-exist [30]. Since there are strong indications that the OP responsible for phase coherence has the magnetic origin [11] there is another interpretation of the tunneling data shown in Fig. 2. Figure 4 shows the case in which the coherent OP consists of the d- and g-wave components. The g-wave component does not change the maximum magnitudes of both gaps, it just adds some "weight" between the maxima of the d-wave and $d_{xy}$ gaps shown in Fig. 4. In this case, both the magnitudes of the two gaps shown in Fig. 4 are slightly different from the similar values presented in Fig. 1. Theoretically, the maximum magnitude of the g-wave gap is about of 30-50% of the maximum magnitude of the d-wave component [30], thus, $\Delta_g \sim$ 6-10 meV. In the case shown in Fig. 4, the pairing OP may have also a strongly anisotropic or extended (turned by 45°) s-wave symmetry. We will analyze all data in Section 6.

## 4. Electronic Raman scattering measurements

Raman measurements do not give the answer to the question what OP has the d-wave symmetry since they show a wide diversity of the symmetries of the two OPs in hole-doped cuprates [18-22,31]. For instance, ERS measurements performed on YBCO and Bi2212 [19] show that the OP responsible for pairing has the d-wave symmetry ($B_{1g}$ polarization). However, in ERS measurements performed on Tl2201 by same group [31], the d-wave symmetry can be attributed to the coherent OP. In ERS measurements on underdoped YBCO and Y-doped Bi2212 [21], the $B_{2g}$ polarization scales with $T_c$ in the underdoped regime. This fact implies that the OP responsible for phase coherence has the $d_{xy}$ symmetry.

From ERS measurements on Bi2212 [18,20] and Tl2201 [20], the OP responsible for phase coherence has a s-wave symmetry ($A_{1g}$ polarization) [18,20]. The OP responsible for pairing has been observed in Bi2212 by the $A_{1g}$, $B_{1g}$ and $B_{2g}$ polarizations [22]. This suggests that the pairing OP can have a s-wave, the d-wave or $d_{xy}$ symmetry. Thus, *both* the OPs have been observed in different hole-doped cuprates by *each* polarization. So, unfortunately, we can not use ERS data to support either APRES or angle-resolved tunneling studies which are at odds with each other. However, in spite of this, we have to admit that, by contrast to ARPES, ERS measurements do show the presence of the two OPs in hole-doped cuprates [18]. This is probably the most important experimental fact from the ERS studies. For example, femtosecond time-domain spectroscopy measurements [32] show also the presence of the two OPs in YBCO. However, it is not the case for ARPES studies.

## 5. The ARPES-tunneling mixed case

Because of the absence of generally accepted model for the ARPES spectral function $A(\mathbf{k}, E)$ in cuprates and very recent Joynt's results [26] presented in Section 3, for the gap *magnitude*, there are more reasons to relay on tunneling measurements than on ARPES data. In spite of this, ARPES technique has definitely one advantage in comparison with tunneling spectroscopy: the angle resolution. Formally, there is no angle resolution in tunneling measurements [27]. At the same time, ARPES technique has an angle resolution of ±1° [23]. Consequently, it is *reasonable* to relay on ARPES *angle*-resolved data keeping the *magnitude* of total gap in accordance with the tunneling results. As we mentioned above, ARPES measurements [17] disagree with the tunneling data [13] by 45°. Thus, by taking into account this disagreement Figure 5 shows an ARPES-tunneling "mixed" case.

By using the new presentation of the tunneling data, shown in Fig. 5, we continue our search of the symmetries of the two OPs. Again, we apply the principle of adjustment and the requirement that one of the two gaps, either the $\Delta_c$ or $\Delta_p$, has the d-wave symmetry. By knowing the magnitudes of the $\Delta_c$ and $\Delta_p$ gaps at $p/p_m$ = 1.2 from Fig. 1 there are a few possible solutions which fit the data in Fig. 5. Figure 6 shows three possible combinations. The first case shown in Fig. 6(a) is definitely the case which corresponds to the ARPES measurements [17] and, probably, the most simplest. There is a good agreement between the gap magnitudes shown in Figs. 1 and 6(a). The second case shown in Fig. 6(b) is also attractive by it's simplicity but the magnitude of the pairing OP is slightly larger than that in Fig. 1. In the case shown in Fig. 6(b), it is possible that the coherent OP has an strongly anisotropic or extended (turned by 45°) s-wave symmetry. The third possible case shown in Fig. 6(c) looks very similar to the $\Delta_p(+)$ solution in Fig. 3(a). In the third case, there is a very good agreement between the gap magnitudes shown in Figs. 1 and 6(c). It is possible that the coherent OP in Fig. 6(c) consists of the d- and g-wave components similar to the case shown in Fig. 4 since the OP responsible for phase coherence has most likely the magnetic origin [11], and the g-wave component does not change the maximum magnitudes of both gaps (see Section 3).

## 6. Discussion

We analyze here the data presented in the previous Sections. We have 11 combinations for the two OPs in hole-doped cuprates: seven cases are presented in Figs. 3, 4 and 6, and there are two additional cases for the data shown in Fig. 6(b), in which the coherent OP may have an anisotropic or extended s-wave symmetry and two additional cases for the data shown in Fig. 4, in which the pairing OP may have an anisotropic or extended s-wave symmetry. We believe that one combination among these 11 cases corresponds to the real case in cuprates. In order to eliminate some combinations we need some experimental data which exclude one or another case.

First of all, we start with in-plane torque anisotropy measurements on Tl2201 [10], which are exclusively sensitive to the magnitude of the OP. It was found that besides the presence of the OP having the d-wave symmetry there exists the second OP which has a s-wave symmetry [10], and they are locked to each other. What is more interesting that the amplitude of the s-wave OP is approximately twice larger than the magnitude of the d-wave component, and the magnitude of the s-wave OP is positive (negative) everywhere [10]. The latter implies that the shape of the s-wave OP is isotropic or anisotropic but not extended. Secondly, there is clear evidence for charge-stripe formation in different cuprates [33-38]. Since $T_{pair} \geq T_c$ [1-4] the pairing occurs above $T_c$, most likely, along charge stripes [39,5]. Recent results show that stripes suppress the d-wave channel of the pair correlation function [37,38]. This implies that the pairing OP has most likely a s-wave symmetry or, at least, the s-wave symmetry is predominant. Thirdly, tunneling measurements on optimally doped Tl2201 show the presence of energy gap with the magnitude of $\Delta$ = 20 - 22 meV and very specific shape of tunneling spectra which can be fit to a simple, momentum-averaged DOS obtained from a SC OP with the d-wave symmetry [3]. Thus, the experimental [10,11,3] and theoretical [37,38] results imply that the OP responsible for phase coherence has the d-wave symmetry, and the pairing OP has the predominant (an)isotropic s-wave symmetry.

Finally, if we accept that the $\Delta_c$ has the magnetic origin [11], then there remains only 3 possible combinations which fit the data: one case is shown in Fig. 3(a), one case is presented in Fig. 6(c), and one combination is shown in Fig. 4 with the pairing OP having an strongly anisotropic s-wave symmetry. In all these three cases, the OP responsible for phase coherence has entirely or partially the d-wave symmetry. The pairing OP has an anisotropic s-wave symmetry. The presence of the g-wave OP responsible for phase coherence, if it exists at all, is probably not universal in all cuprates [5] and depends on hole concentration [30,40]. The s-wave symmetry of the pairing OP can explain the s-wave character of the SC in electron-doped $Nd_{2-x}Ce_xCuO_4$ (NCCO) cuprate [41]. The s-wave symmetry of total OP in NCCO implies that the OP responsible for phase coherence in NCCO has a s-wave symmetry, consequently, it has the non-magnetic origin [30]. Indeed, many physical properties of NCCO are different from hole-doped cuprates [42,43,8,5].

One can argue that our conclusions contradict to ARPES measurements. Indeed, the ARPES case implies that the pairing OP has the d-wave symmetry. However, we find that many experimental data which are at odds with ARPES measurements are consistent among themselves and they are much reliable than ARPES data [26]. Even, if ARPES data are correct, they reflect the DOS of the quasiparticle excitations in the thin layer of 3 Å on the surface [25] whereas, for example, the torque measurement is a true *bulk* experiment [10].

### 7. Conclusions

In summary, we discussed here the symmetries of the two OPs in hole-doped cuprates. Our analysis based mainly on tunneling and torque measurements shows that, most likely, the OP responsible for phase coherence has the $d_{x^2-y^2}$ symmetry whereas the OP responsible for pairing has an anisotropic s-wave symmetry. By taking into account that the origin of the OP responsible for phase coherence is most likely magnetic it is possible that the coherent OP *may* consists of the $d_{x^2-y^2}$ and g-wave components.

**Acknowledgements**

The author thanks R. Deltour for discussion. This work is supported by PAI 4/10.

FIGURE CAPTIONS:

Fig. 1. Phase diagram in hole-doped cuprates: $\Delta_c$ is the coherent OP, and $\Delta_p$ is the pairing OP [1]. The $p_m$ is a hole concentration with the maximum $T_c$.

Fig. 2. Tunneling gap at low temperature vs. angle: dots (average measured points) and solid line (an assumption of a fourfold symmetry) [13]. The d-wave gap is shown schematically with the maximum magnitude of 20 meV (see text).

Fig. 3. Shapes of two OPs at low temperature in overdoped Bi2212: (a) the d-wave coherent OP and an anisotropic (+) or "extended" (-) s-wave pairing OP, and (b) the d-wave coherent OP and the pairing OP mixed of the isotropic s-wave and the $d_{xy}$ components. The shapes of OPs are shown schematically.

Fig. 4. Shapes of two OPs at low temperature in overdoped Bi2212: the coherent OP mixed of the d- and g-wave components and the $d_{xy}$ pairing OP. The shapes of OPs are shown schematically. The unmarked lobes of the g-wave OP have the positive sign. The pairing OP can have also either an anisotropic or "extended" s-wave symmetry (see text).

Fig. 5. The tunneling data from in Fig. 2 turned by 45° (see text).

Fig. 6. (a), (b) and (c) Different combinations of shapes of the two OPs at low temperature in overdoped Bi2212 which fit the data in Fig. 5. The coherent OP in the case (b) can have also either an anisotropic or "extended" s-wave symmetry. The shapes of OPs are shown schematically.

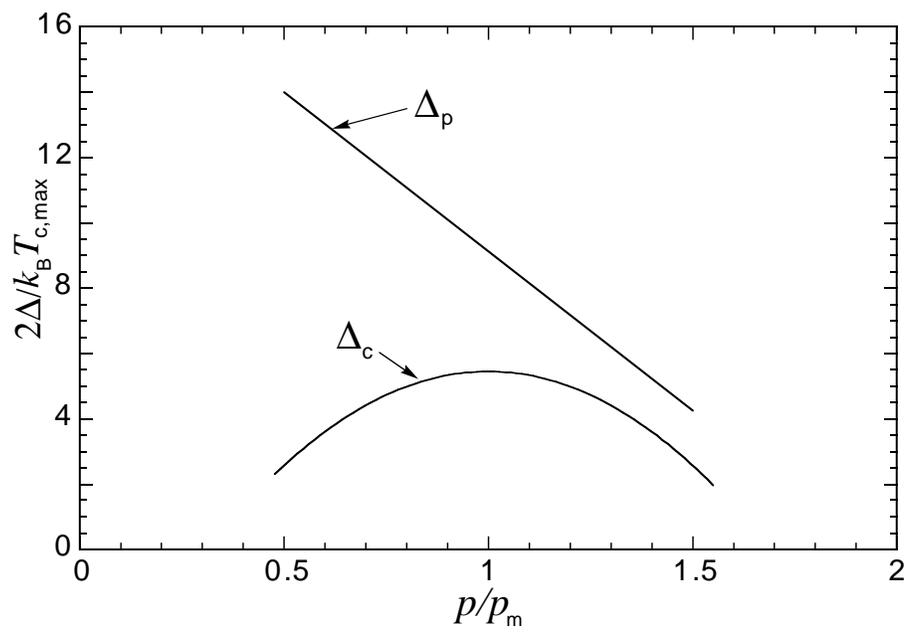

FIG. 1

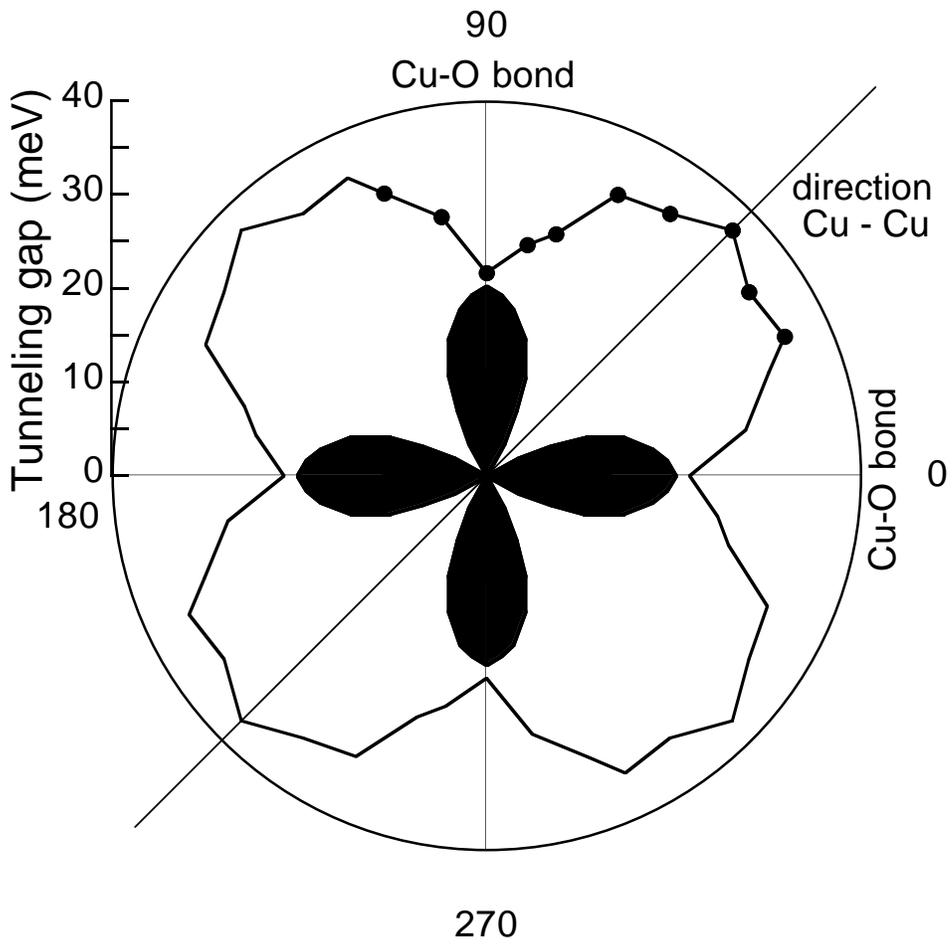

FIG. 2

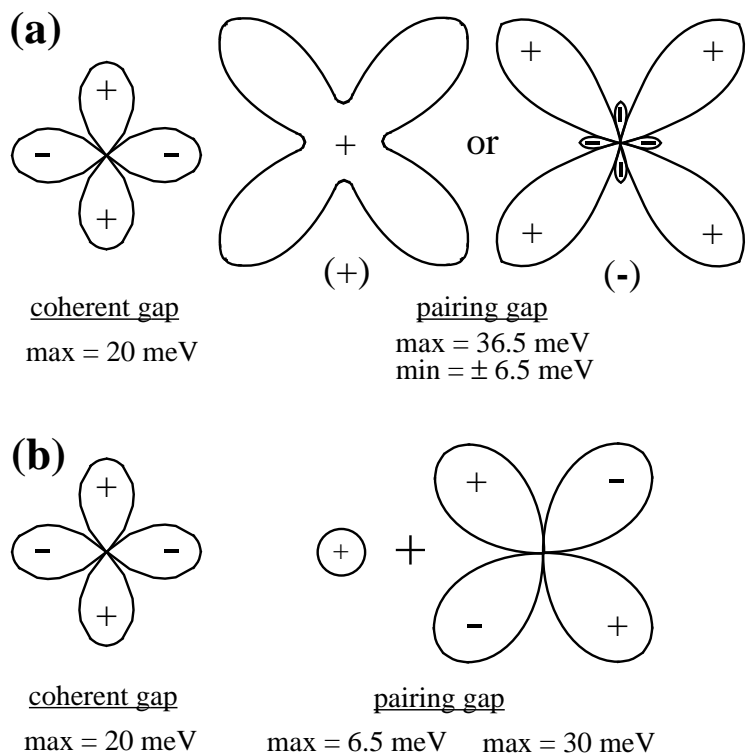

FIG. 3

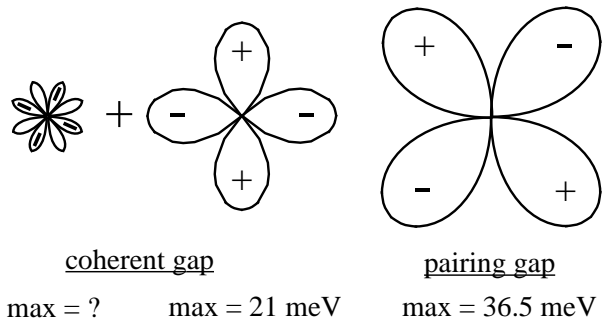

coherent gap      pairing gap

max = ?     max = 21 meV     max = 36.5 meV

FIG. 4

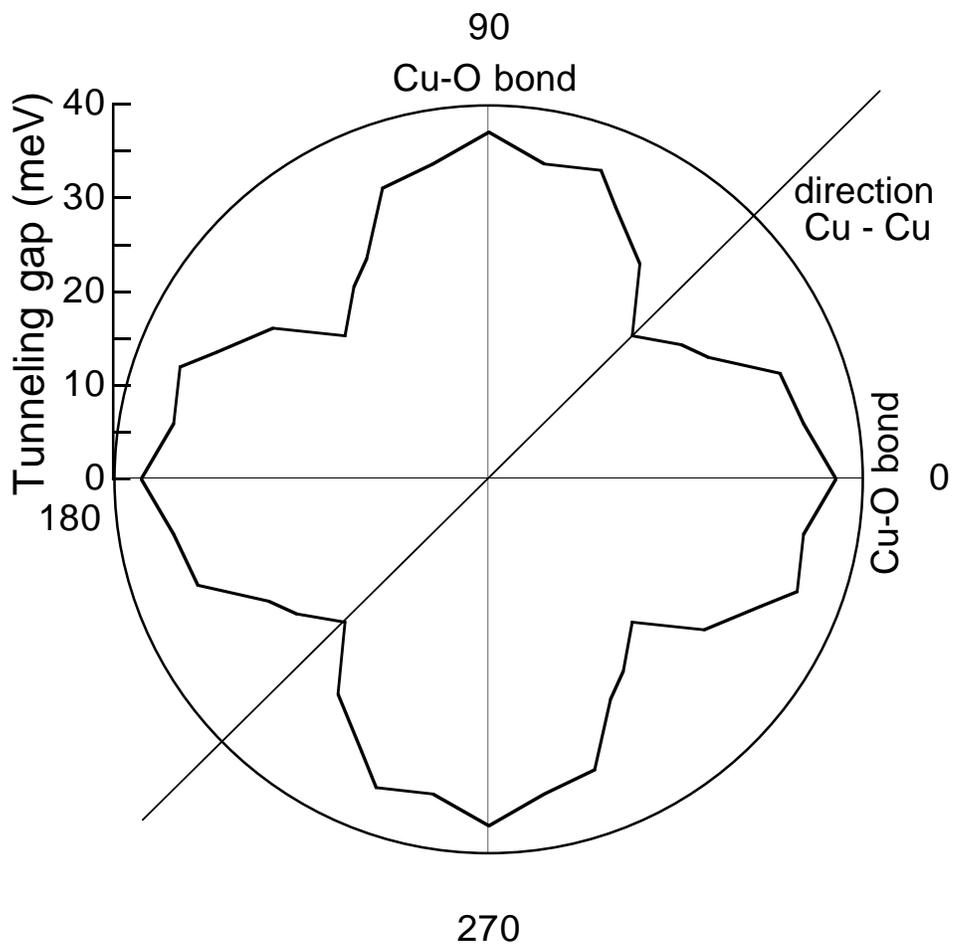

FIG. 5

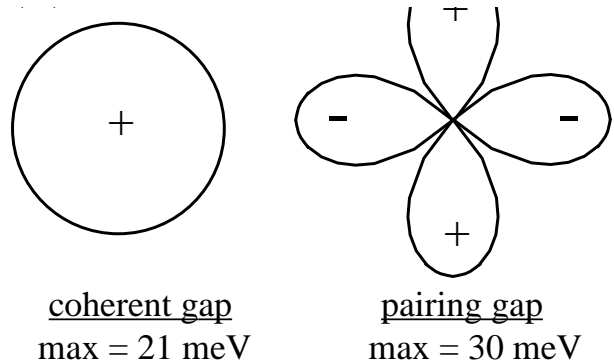

coherent gap  
max = 21 meV

pairing gap  
max = 30 meV

**(b)**

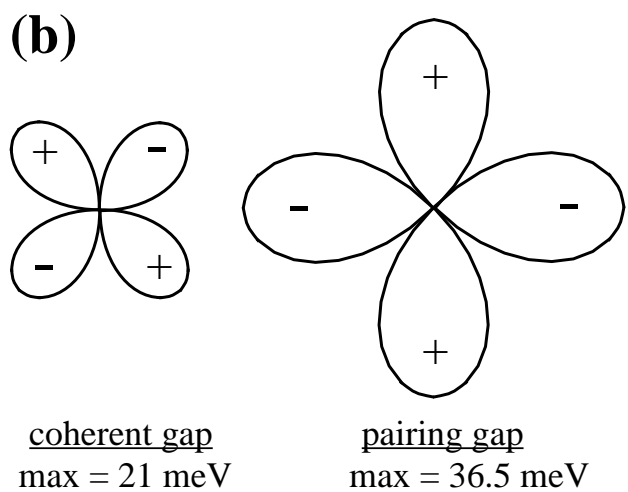

coherent gap  
max = 21 meV

pairing gap  
max = 36.5 meV

**(c)**

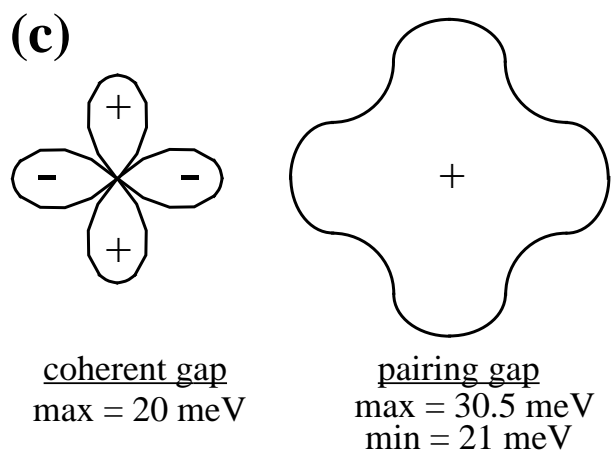

coherent gap  
max = 20 meV

pairing gap  
max = 30.5 meV  
min = 21 meV

FIG. 6